%% file: arxiv.tex
\documentclass[sigconf,natbib=true]{acmart}
\AtBeginDocument{%
  \providecommand\BibTeX{{%
    \normalfont B\kern-0.5em{\scshape i\kern-0.25em b}\kern-0.8em\TeX}}}

\usepackage{bm}
\usepackage{enumitem}
\usepackage{graphicx}
\usepackage{amsthm}
\usepackage{sidecap}
\usepackage{mathtools}
\usepackage{enumitem}
\usepackage{titlesec}
\usepackage{caption}
\usepackage{hyperref}
\usepackage{xcolor}

\newtheoremstyle{mystyle} 
    {0pt}                
    {0pt}                
    {\normalfont}        
    {}                   
    {\bfseries}          
    {.}                  
    { }                  
    {}                   

\theoremstyle{mystyle}
\newtheorem{definition}{Definition}

\newcommand{\header}[1]{\noindent \textbf{#1}}
\newcommand{\modelname}{PromptLink}
\newcommand{\datasetone}{MIID}
\newcommand{\datasettwo}{CISE}

\newcommand{\humanEmoji}{\includegraphics[height=1em]{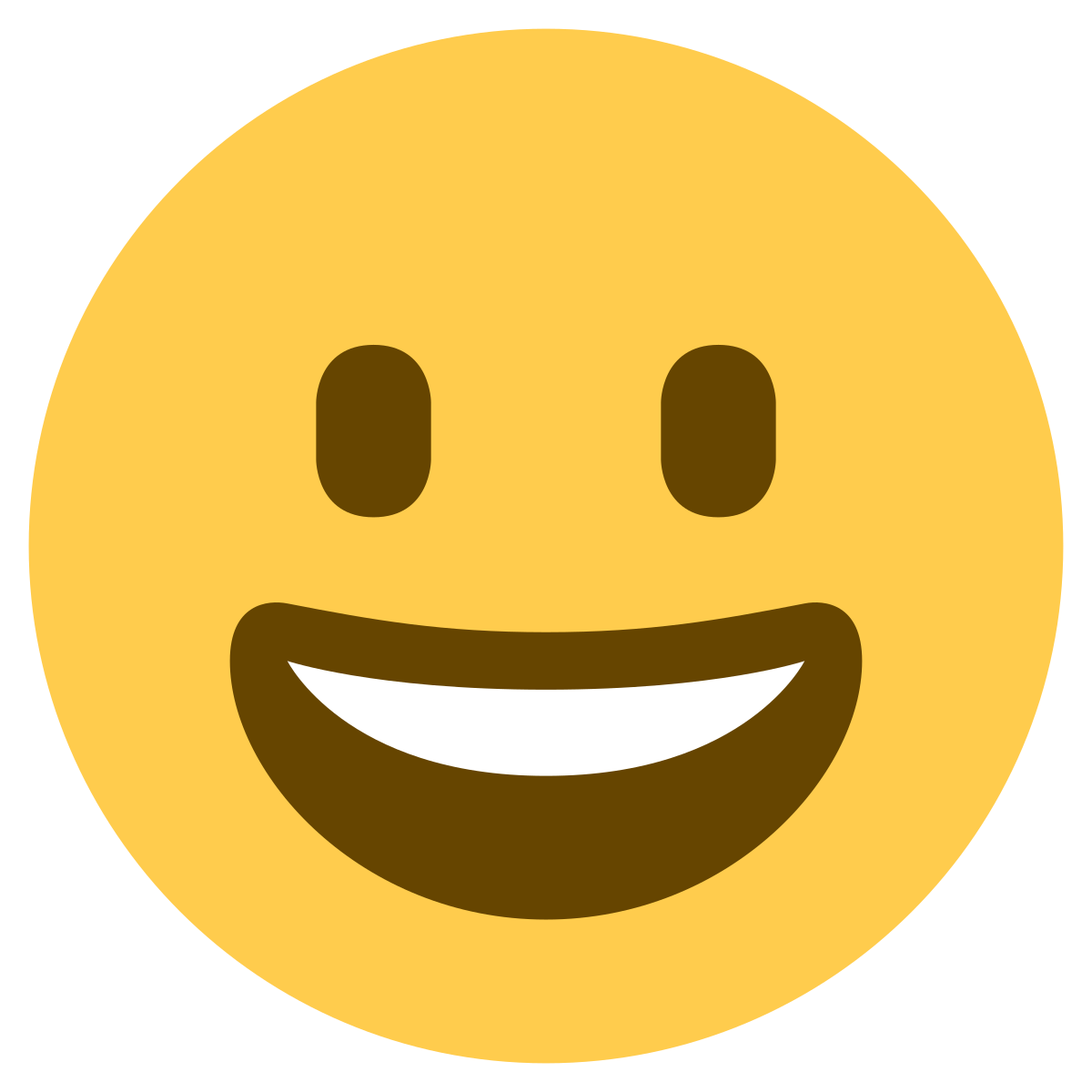}}
\newcommand{\labelEmoji}{\includegraphics[height=1em]{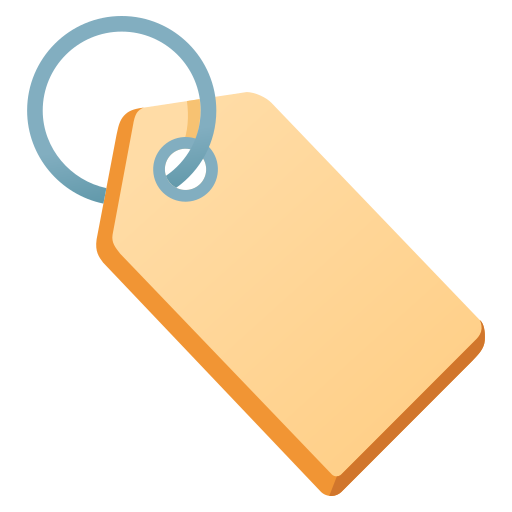}}

\input{sections/1_format}

\begin{document}

\input{sections/2_abstract}

\title{\modelname: Leveraging Large Language Models for Cross-Source Biomedical Concept Linking}
\maketitle

\input{sections/3_intro}

\input{sections/4_method}
\input{sections/5_experiment}
\input{sections/6_conclusion}
\appendix
\vspace{3pt}
\section*{Acknowledgements}
Research reported in this publication was partially supported by the National Institute Of Diabetes And Digestive And Kidney Diseases of the National Institutes of Health under Award Number K25DK135913. JH was supported by the National Science Foundation under Award Number IIS-2145411.
\newpage
\bibliographystyle{ACM-Reference-Format}
\bibliography{A}
\end{document}

%% file: sections/1_format.tex
\copyrightyear{2024}
\acmYear{2024}
\setcopyright{acmlicensed}
\acmConference[SIGIR '24] {Proceedings of the 47th International ACM SIGIR Conference on Research and Development in Information Retrieval}{July 14--18, 2024}{Washington, DC, USA.}
\acmBooktitle{Proceedings of the 47th International ACM SIGIR Conference on Research and Development in Information Retrieval (SIGIR '24), July 14--18, 2024, Washington, DC, USA}
\acmISBN{979-8-4007-0431-4/24/07}
\acmDOI{10.1145/3626772.3657904}

\author{Yuzhang Xie}
\affiliation{Emory University\country{USA}}
\email{yuzhang.xie@emory.edu}

\author{Jiaying Lu}
\affiliation{Emory University\country{USA}}
\email{jiaying.lu@emory.edu}

\author{Joyce Ho}
\affiliation{Emory University\country{USA}}
\email{joyce.c.ho@emory.edu}

\author{Fadi Nahab}
\affiliation{Emory University\country{USA}}
\email{fnahab@emory.edu}

\author{Xiao Hu}
\affiliation{Emory University\country{USA}}
\email{xiao.hu@emory.edu}

\author{Carl Yang}
\affiliation{Emory University\country{USA}}
\email{j.carlyang@emory.edu}

\renewcommand{\shortauthors}{Yuzhang Xie et al}


%% file: sections/2_abstract.tex
\begin{abstract}
Linking (aligning) biomedical concepts across diverse data sources enables various integrative analyses, but it is challenging due to the discrepancies in concept naming conventions. 
Various strategies have been developed to overcome this challenge, such as those based on string-matching rules, manually crafted thesauri, and machine learning models. However, these methods are constrained by limited prior biomedical knowledge and can hardly generalize beyond the limited amounts of rules, thesauri, or training samples. 
Recently, large language models (LLMs) have exhibited impressive results in diverse biomedical NLP tasks due to their unprecedentedly rich prior knowledge and strong zero-shot prediction abilities. However, LLMs suffer from issues including high costs, limited context length, and unreliable predictions. 
In this research, we propose \modelname, a novel biomedical concept linking framework that leverages LLMs. It first employs a biomedical-specialized pre-trained language model to generate candidate concepts that can fit in the LLM context windows. Then it utilizes an LLM to link concepts through two-stage prompts, where the first-stage prompt aims to elicit the biomedical prior knowledge from the LLM for the concept linking task and 
the second-stage prompt enforces the LLM to reflect on its own predictions to further enhance their reliability. 
Empirical results on the concept linking task between two EHR datasets and an external biomedical KG demonstrate the effectiveness of \modelname. 
Furthermore, \modelname\ is a generic framework without reliance on additional prior knowledge, context, or training data, making it well-suited for concept linking across various types of data sources. 
The source code of this study is available at \href{https://github.com/constantjxyz/PromptLink}{https://github.com/constantjxyz/PromptLink}.
\end{abstract}

\ccsdesc[500]{Applied computing}
\ccsdesc[500]{Applied computing~Health care information systems}
\ccsdesc[500]{Information systems}
\ccsdesc[300]{Information systems~Retrieval models and ranking}

\keywords{Biomedical Concept Linking, Few-Shot Prompting, Large Language Models for Resource-Constrained Field, Retrieve \& Re-Rank}

%% file: sections/3_intro.tex
\section{Introduction}

\begin{figure}[htbp]
\centering
\vspace{2pt}
\includegraphics[width=1.0\linewidth]{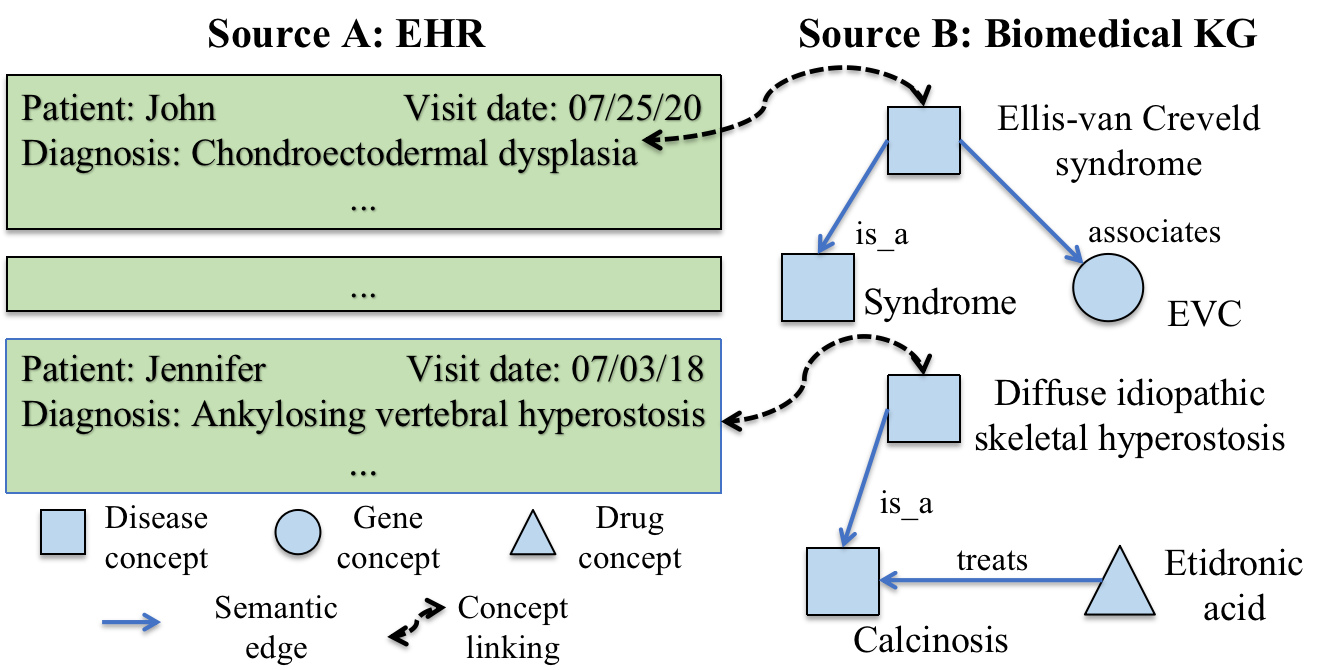}
\caption{A toy example of biomedical concept linking. \textit{Left:} concepts in the EHR. \textit{Right:} concepts in the biomedical KG.}
\label{fig:intro}
\vspace{1pt}
\end{figure}

Biomedical concept linking studies the intricate task of linking closely related concepts across different data sources by leveraging their semantic meanings and underlying biomedical knowledge, as exemplified in Figure \ref{fig:intro} \cite{sevgili2022EL}.
This linking process is crucial for enabling integrative analyses, as biomedical concepts obtained from diverse sources offer multifaceted views of biomedical knowledge and data~\cite{su2023ibkh,lu2023hiprompt}. For example, the electronic health record (EHR), which is regarded as a valuable asset for comprehensive patient health analysis, contains various digital medical information including tabular data, clinical notes, and other types of patient data \cite{abul2019EHR, sun2018EHR, xu2022CACHE}. Similarly, the knowledge graph (KG), playing an important role in biomedical research, provides structured knowledge, such as definitions of concepts and their interrelationships \cite{ma2018EHRKG}. 
However, the cross-source biomedical linking task is challenging due to discrepancies in the biomedical naming conventions used in different systems \cite{kohane2021EHRKG}. For example, a KG may mention a disease as ``Ellis-Van Creveld syndrome'', while an EHR may refer to the same disease as ``Chondroectodermal dysplasia''. This inconsistency presents a strong barrier to cohesive data analysis.

The challenge of biomedical concept linking has motivated the development of various methods. \textbf{Conventional methods} focus on setting \textit{string-matching rules} \cite{d2015ELRule,kang2013ELRule} and leveraging \textit{constructed thesauri} \cite{aronson2010ELDICT,savova2010ELDICT,friedman2001ELDICT}. However, their reliance on fixed rules and crafted thesauri limits coverage and generalizability in real-world scenarios \cite{shi2023EL}.
Addressing these limitations, \textbf{machine learning-based methods} have been widely explored, 
avoiding the manual design of rules or thesauri. These methods essentially transform biomedical concepts from raw text into embeddings (latent vector representations), which are then used to compute similarity scores via distance functions (\textit{e.g.} cosine similarity) or learning-based scoring functions (\textit{e.g.} bilinear attention~\cite{kim2018bilinear}). 
Various models have been used to obtain biomedical concept embeddings, including \textit{pre-trained language models}  (PLMs)~\cite{wang2023ELLM} that capture fine-grained semantic relations through extensive training on biomedical corpora ~\cite{xu2020ELLM,lee2020biobert, alsentzer2019bioclinicalbert, liu2020sapbert}, and \textit{graph neural networks} (GNNs)~\cite{zhou2020ELMLGNN} that capture both semantics and relations of biomedical concepts \cite{bordes2013ELML,grover2016ELML,liu2022ELML}.
Despite the notable achievements of these ML-based linking methods, they are data-hungry and require significant supervision signals when adapted into novel downstream applications. They face challenges due to the costly data annotation and model training processes.


Recently, large language models (LLMs) have exhibited impressive performances in various NLP tasks, due to their unprecedentedly rich prior knowledge and language capabilities \cite{zhou2023LLM, singhal2022LLM, wang2023LLM}, enabling various applications in a zero-shot learning setting \cite{lu2023hiprompt}. 
Therefore, LLMs provide a promising solution for linking related concepts across different systems.
Meanwhile, LLMs also face challenges including the design of effective and cost-efficient prompts within the context length limits \cite{zhang2023LLM}, and the NIL prediction capability of reliably rejecting all candidates when correct concepts are absent, instead of returning relatively close but incorrect ones \cite{peters2019NIL}.

In this paper, we propose \textbf{\modelname}, leveraging LLMs for the cross-source biomedical concept linking task. Considering LLMs' high cost and context length constraints, we first employ a pre-trained SAPBERT language model to generate biomedical-aware concept embeddings and retrieve top candidates based on the cosine similarities of these embeddings. We then design a novel two-stage prompting mechanism for the GPT-4 model to derive reliable linking predictions. The first stage efficiently filters out irrelevant candidates, thereby minimizing the response token numbers required in the subsequent stage. The second stage generates the final linking results and incorporates a self-verification prompt to address the NIL prediction challenge, effectively rejecting all candidates when none are relevant. In the experiments, \modelname\ demonstrates exceptional performance, surpassing various existing concept linking methods by over 5\% in two scenarios of biomedical concept linking between EHR and external biomedical KG, which could be attributed to LLM's intrinsic strong biomedical knowledge. Moreover, \modelname\ works as a zero-shot framework due to the utilization of pre-trained language models, eliminating the need for a training process. It is also a versatile framework that performs well even when only concept names, without concept context or topological structure, are provided. Given its various advantages, \modelname\ boasts strong generalization capabilities, making it suitable for a wide range of biomedical research and applications.

%% file: sections/4_method.tex
\setlength{\abovedisplayskip}{0pt} 
\setlength{\belowdisplayskip}{0pt}

\section{Biomedical Concept Linking}
\begin{figure*}[htbp]
\centering
\includegraphics[width=0.8\linewidth]{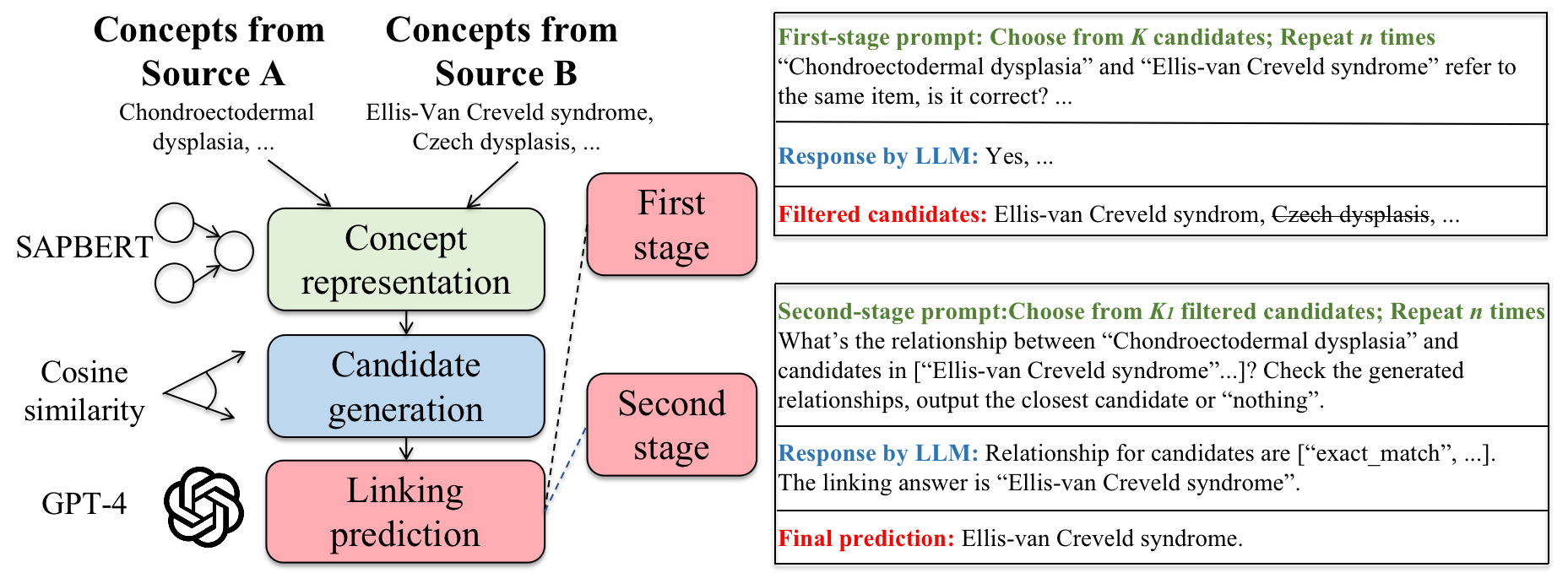}
\caption{Overview of our proposed \modelname\ framework.}
\label{fig:method}
\vspace{-10pt}
\end{figure*}

\subsection{Problem Definition}
\label{ssec:problem_def}
The biomedical concept linking task aims to link biomedical concepts across sources/systems based on semantic meanings and biomedical knowledge. It solely relies on concept names and can thus cover much broader real-world applications. This task differs from existing tasks such as entity linking~\cite{shi2023EL}, entity alignment~\cite{lu2023hiprompt}, and ontology matching~\cite{jimenez2011logmap}, which depend on extra contextual or topological information. 
In this study, we link the concepts in EHR to corresponding concepts in a biomedical KG. 
We define an EHR database $\mathcal{D}$, a biomedical KG $\mathcal{G}$, and the linking task as follows:
\begin{definition}[EHR]
An EHR database $\mathcal{D}$ is a relational database $\mathcal{D}=(P,A,V)$, with $P$ being patient identifiers,  $A$ patient attributes, $V\in P\times A$ the values of these attributes. Additionally, $M$ represents multi-token biomedical concepts associated with patient attributes.
\end{definition}
\begin{definition}[Biomedical KG]
A biomedical KG is a multi-relation graph $\mathcal{G}=(C,R,RT)$, where $C$ are concepts, $R$ are relation names, and $RT\in C\times R \times C$ are the relational triples among them.
\end{definition}
\begin{definition}[Biomedical Concept Linking]
Link identified biomedical concepts from an EHR $\mathcal{D}$ to a biomedical KG $\mathcal{G}$ based on semantic meanings and biomedical knowledge, forming linkages $LK=\{(m_i, c_j) | m_i\in M_{\mathcal{D}},c_j\in C_{\mathcal{G}}\cup \text{NIL}\}$. If a concept $m$ from $D$ is not in $G$, link it to a special ``\textit{NIL}'' entity, indicating it is unlinkable.
\end{definition}

\subsection{\modelname}
We propose \modelname, a novel LLM-based solution for cross-source biomedical concept linking, as illustrated in Figure \ref{fig:method}.
Addressing LLMs' high cost and limited input text length, we first employ a biomedical-specialized pre-trained language model 
to generate concept embeddings and retrieve top candidates via cosine similarities. Subsequently, we employ a two-stage prompting mechanism with GPT-4 to generate the final linking predictions.

\header{Concept representation and candidate generation.}
After preprocessing text by lowercasing and removing punctuation, we use a pre-trained LM (specifically \textit{SapBERT}~\cite{liu2020sapbert}), 
to create embeddings $\bm{h} \in \mathbb{R}^{1\times d}$ for EHR concepts $m$ and KG concepts $c$, represented as 
\vspace{2pt}
\begin{equation*}
    \bm{h_m} = PLM(m), \quad \bm{h_c} = PLM(c).
\end{equation*}
For concepts that span multiple tokens, the token-level embeddings are averaged to create the concept embedding.
This model helps to project the semantic meanings and prior biomedical knowledge into the embedding space.
For candidate generation, we compute cosine similarity $S \in [0, 1]$ between pairs of EHR concept embedding $\bm{h_m}$ and KG concept embedding $\bm{h_c}$, represented as 
\begin{equation*}
     S = cos( \bm{h_{m}}, \bm{h_{c}} ).
\end{equation*}
Given each input query EHR concept $m_i$, We select the top-$K$ ($K$=10) KG concepts $[c_1, c_2, \dots, c_K]$ with the highest similarities as candidates for further GPT-based linking prediction.

\vspace{2pt}
\header{Linking prediction using two-stage prompts.}
The next step of our framework is generating linking predictions of query $m_i$ over the top-$K$ candidate $[c_1, c_2, \dots, c_K]$ using GPT-4 model, leveraging its text comprehension ability, logical reasoning ability, and prior biomedical knowledge \cite{bubeck2023LLM, singhal2022LLM}. In this step, we design a novel two-stage prompt for our task, as can be seen in Figure~\ref{fig:method}.
\noindent Combining the two prompts utilizes their strengths and mitigates weaknesses. The first stage focuses on concept pairs to filter out unrelated candidates. The second stage evaluates all candidates in a broader context to identify the closest match or reject all unmatch candidates.


In the first stage, the LLM is prompted to check if a concept pair $(m_i, c_j)$ should be linked. By defining the response structure, the LLM can return answers in specified formats. To improve the prompt response quality, we adopt the self-consistency~\cite{wang2022selfcons} prompting strategy that repeatedly prompts the same question to the LLM multiple times.
Specifically, we prompt each concept pair $(m_i, c_j)$ for $n=5$ times, thus obtaining the belief score $B_{i,j} \in [0, 1]$ by:
\vspace{3pt}
\begin{equation*}
    B_{i,j} = \frac{\text{number\ of\ ``yes'' responses}}{n}.
\vspace{4pt}
\end{equation*}
Considering the belief scores across different candidates, we derive a comprehensive filter strategy to exclude irrelevant candidates, using parameter $\tau$ (set as $0.8\times n$). This approach ensures that irrelevant candidates are not considered in the next stage, optimizing both efficiency and effectiveness. The approach is described as follows:
\begin{itemize}[leftmargin=*, noitemsep, topsep=0pt, parsep=0pt, partopsep=0pt]
    \item If $max(B_{i,1}, \cdots, B_{i,K}) \geq \tau$, this indicates some candidates closely align with the query concept. In such cases, candidates with belief scores of zero will be filtered out as they are deemed irrelevant to the query concept and there are closely aligning alternatives. This filtering strategy effectively removes many irrelevant candidates, thereby optimizing both efficiency and effectiveness for the subsequent stage.
    \item Otherwise, the range of different candidates' belief scores is not wide enough to justify filtering. Thus, all $K$ candidates will be subjected to double-checking by the second-stage prompt.
\end{itemize}


In the second stage, the LLM evaluates the $K_1$ candidates retained from the first stage's filtering process $[c'_1,c'_2,\dots,c'_{K_1}]$, where $K_1 \leq K$ , using a compositional prompt that consists of two consecutive questions to perform complex reasoning.
Specifically, the LLM is asked to (1) label the relationship between the query concept and all candidate concepts as ``exact match'', ``related to'', or ``different from''; (2) use self-verification prompts to either identify the closest candidate or dismiss all candidates if none are close, thus the final concept linking result of this prompt is $K_2$ (usually $K_2=1$) item from $[c'_1,\dots,c'_{K_1}] \cup  [\text{NIL}]$. 
In this stage, we also use the self-consistency strategy that prompts one question for the same $n$ times. Subsequently, we calculate the occurrence frequency $f_{i,j} \in [0, 1]$ for answers in $[c'_1,c'_2,\dots, \text{NIL}]$ and retrieve the final linking result for query EHR concept $m_i$ as follows:
\begin{itemize}[leftmargin=*, noitemsep, topsep=0pt, parsep=0pt, partopsep=0pt]
    \item If $f_{i,\ \text{NIL}} > 0.5$, this indicates a high probability that none of the candidates are appropriate. Thus, ``NIL'' is chosen as the final linking prediction.
    \item Otherwise, the candidate $c_j$ with the highest frequency $f_{i, j}$ is decided as the final linking result. If two candidates tie for the highest frequency, the one $c_j$ with higher embedding similarity $S_{i,j}$ to the query concept $m_i$ is chosen.
\end{itemize}

%% file: sections/5_experiment.tex
\vspace{3pt}
\section{Experiments \& Discussions}
\subsection{Implementation Details}
\header{Datasets.}
In our experiments, we curate two biomedical concept linking benchmark datasets:  \textit{\datasetone} (\textbf{M}IMIC-\textbf{I}II-\textbf{i}BKH-\textbf{D}isease) and \textit{\datasettwo} (\textbf{C}RADLE-\textbf{i}BKH-\textbf{S}ide-\textbf{E}ffect). 
\datasetone\ comprises 1,493 diagnosis concepts from MIMIC-III~\cite{johnson2016mimic}, which is an EHR dataset including over 53,423 hospital patient records, and 18,697 disease concepts from iBKH~\cite{su2023ibkh}, which is a KG dataset with 2,384,501 entities. To construct \datasetone, we first remove exact matches between MIMIC-III diagnosis concepts and iBKH disease concepts. Then, we link the remaining MIMIC concepts to iBKH using  ICD-9~\cite{CDC_ICD9CM} and UMLS CUI~\cite{schuyler1993umls} codes. We use the linked concept pairs as ground-truth labels only for evaluation purposes.
\datasettwo\ contains 1,500 CRADLE~\cite{xu2022CACHE} diagnosis concepts and 4,251 iBKH drug side-effect concepts, constructed by using CUI~\cite{schuyler1993umls} and SNOMED CT~\cite{donnelly2006snomedct} codes. Ground-truth matched pairs are also only used for evaluation purposes.


\header{Experimental Settings.}
Following the definition in Sec.~\ref{ssec:problem_def} and recognizing the scarcity of supervision in the biomedical domain, we mainly focus on the biomedical concept linking under the zero-shot setting. 
Additionally, our biomedical concept linking task solely relies on concept names for broad real-world application coverage. 
Given this characteristic of our data input, graph-based linking methods, such as selfKG~\cite{liu2022ELML}, are not applicable as they need topological information to establish concept alignment.
Similarly, thesauri-based methods, such as MetaMap~\cite{aronson2010ELDICT}, are unsuitable as they only establish links between EHR concepts and KG concepts existing in the pre-defined vocabulary. 
Therefore, the following baseline methods are compared: 
\begin{itemize}[leftmargin=*, noitemsep, topsep=0pt, parsep=0pt, partopsep=0pt]
    \item Conventional methods: \textbf{Cosine Distance}, \textbf{Jaccard Distance}, \textbf{Levenshtein Distance}~\cite{ristad1998EditDistance}, \textbf{Jaro-Winkler Distance}~\cite{winkler1990JWDistance}, \textbf{BM25} \cite{robertson2009BW25}. These methods measure the concept pairs' string similarity and relevance and then obtain the linking prediction result.
    \item Machine learning-based methods: Pre-trained language models are used to generate concept embedding and linking prediction results (according to embedding cosine similarity). Specifically, we select representative PLMs including \textbf{BioBERT}~\cite{lee2020biobert}, \textbf{BioGPT}~\cite{luo2022biogpt}, \textbf{BioClinicalBERT}~\cite{alsentzer2019bioclinicalbert}, \textbf{BioDistilBERT}~\cite{rohanian2023biodistilbert}, \textbf{KRISSBERT}~\cite{zhang2022krissbert}, \textbf{ada002}~\cite{neelakantan2022ada002}, and \textbf{SAPBERT}~\cite{liu2020sapbert}.
\end{itemize}

\vspace{1pt}
\subsection{Concept Linking Experiment Results}
\begin{table}[h!]
    \caption{Comparison of the zero-shot accuracy for different methods on \datasetone\ and \datasettwo.}
    \label{tab:acc}
    \centering
    \begin{tabular}{ccc}
    \hline
    Method  & Acc-\datasetone & Acc-\datasettwo \\
    \hline
    Cosine Distance & 0.2981& 0.2907\\
    Jaccard Distance & 0.2123& 0.3280\\
    Levenshtein Distance & 0.1995 & 0.3033\\
    Jaro-Winkler Distance & 0.3141 & 0.3693 \\
    BM25 & 0.4722 & 0.3993\\
    \hline
    BioBERT & 0.3423 & 0.5280 \\
    BioClinicalBERT & 0.3007 & 0.5007\\
    BioGPT & 0.3530 & 0.5093 \\
    BioDistilBERT & 0.4240 & 0.5293 \\
    KRISSBERT & 0.5265& 0.5787\\
    ada002 & 0.5968 & 0.6773 \\
    SAPBERT& 0.7213 & 0.8167 \\
    \hline
    \modelname & \textbf{0.7756} & \textbf{0.8880} \\
    \hline
    \end{tabular}
\end{table}


Table~\ref{tab:acc} shows the accuracy of our proposed \modelname\ along with baseline methods, when every method links a query EHR concept $m_i$ with their predicted top-1 KG concept $c_j$.
As can be seen, \modelname\ outperforms competing approaches across both datasets in terms of zero-shot accuracy, underscoring the superiority of our LLM-based concept linking methodology. 
Among the compared methods, SAPBERT, a SOTA biomedical entity linking method, achieves the second-highest performance. 
Moreover, conventional methods based on string similarity lag behind machine learning techniques, which leverage embeddings from pre-trained language models to effectively match conceptually similar but lexically distinct entities like ``Ellis-Van Creveld syndrome'' and ``Chondroectodermal dysplasia''.

\subsection{Ablation Studies}
\begin{table}[htbp!]
    \caption{Ablation results with different prompting methods used by \modelname\ on the \datasetone\ dataset.}
    \label{tab:prompt}
    \centering
    \begin{tabular}{ccc}
    \hline
       Prompting Methods  & Acc & Token Cost \\
    \hline
       Before Prompting & 0.7213 & N/A \\
       First-stage Prompt & 0.7595 & 995,836 (\$36.59)\\
       Second-stage Prompt & 0.7634 & 1,681,987 (\$88.69)\\
       Two-stage Prompts & \textbf{0.7756} & 1,594,996 (\$66.25) \\
    \hline
    \end{tabular}
\end{table}

\vspace{3pt}
\header{Prompt Effectiveness and Efficiency.}
We conduct ablation studies to reveal the effectiveness and cost-efficiency of the prompt used in our approach, as shown in Table \ref{tab:prompt}. This comparison uses the same input data and 10 linking candidates across various prompts.  In the table, the ``Before prompting'' denotes the performance of using only embedding similarity obtained from the pre-trained LM, while other methods use LLM to predict linking results based on LM-generated candidates. 
From Table \ref{tab:prompt}, the ``Before Prompting'' method achieves the worst accuracy, demonstrating that linking performance could be improved by using LLM. 
Notably, \modelname with both two-stage prompts achieves the best accuracy with the second-highest cost ($\sim1.7$M total tokens, costing approximately \$66.25), indicating that the combined effect of the prompts substantially enhances accuracy, with the costs being moderated by the first stage's proficiency in eliminating unrelated candidates.

\begin{SCfigure}[][htbp]
    \centering
    \vspace{-2pt}
    \includegraphics[width=0.5\linewidth]{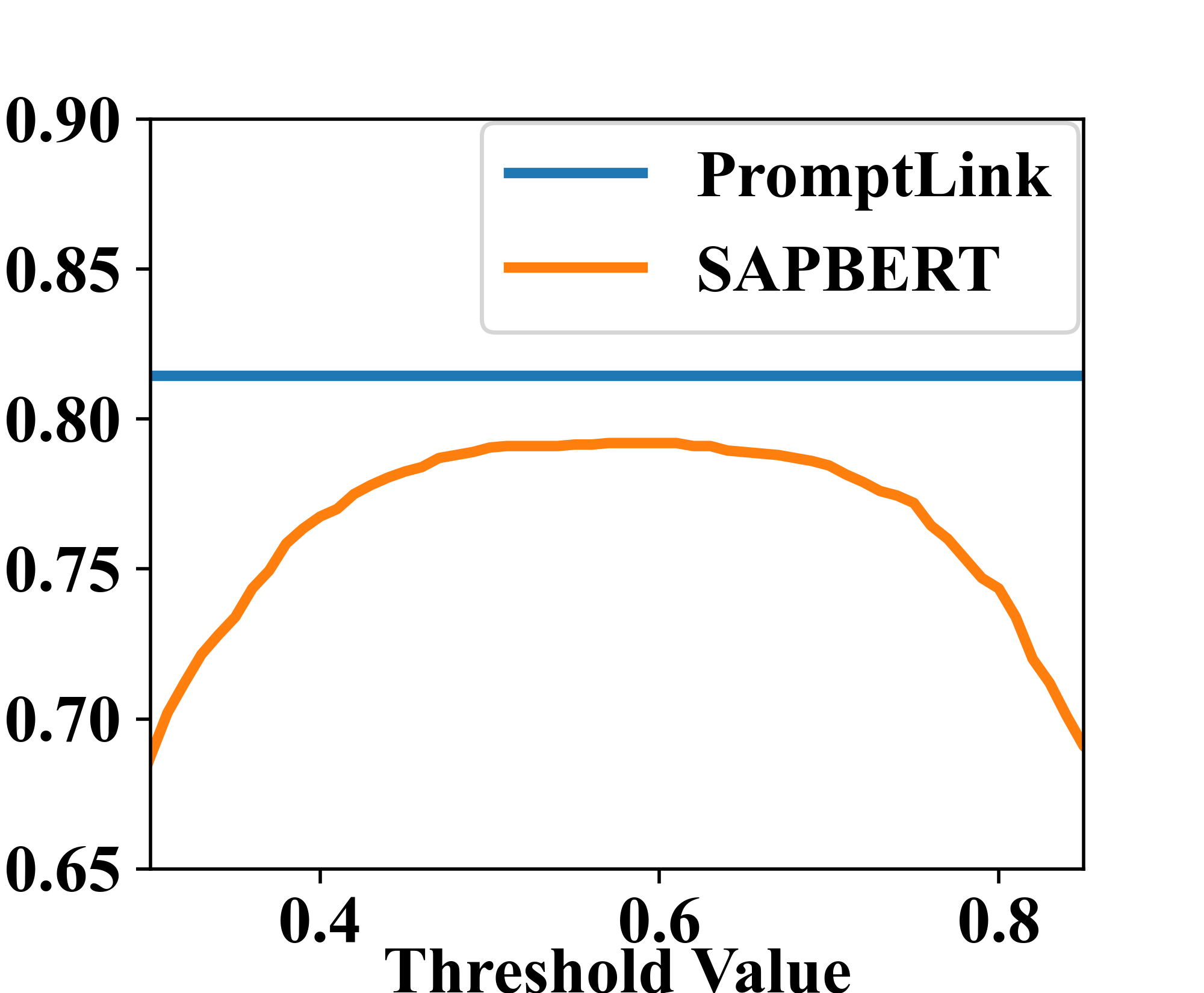}
    \caption{Accuracy on MIID-NIL: \textnormal{Traditional ML-based methods outputting matching scores have varying NIL prediction performance based on the selected threshold, while \modelname\ does not need a threshold yet consistently performs better.}}
    \label{fig:nil}
\end{SCfigure}

\vspace{2pt}
\header{NIL Prediction.}
Another ablation study examines \modelname's NIL prediction ability. 
In our built MIID and CISE datasets, each query EHR concept $m_i$ is designed to have a ground-truth linking KG concept $c_j$. To reflect the real-world unlinkable scenario, we extend our MIID dataset into ``\textit{MIID-NIL}'' which contains a proportion ($25\%$) of unlikable EHR concept $m_i$.
In Figure \ref{fig:nil}, the overall accuracy of \modelname\ in the MIID-NIL dataset is 0.8145. 
Specifically for the unlikable concepts, \modelname\ outputs the expected ``NIL'' with 0.9290 accuracy, which validates the NIL prediction ability of our proposed method. 
Existing methods highly rely on the hard-coded threshold. For example, we could threshold SAPBERT's generated embeddings' cosine similarity, then output the KG concept with the highest similarity above the threshold or ``NIL'' when none are above. However, this straightforward idea, requiring a manually set threshold, is less effective than \modelname. As shown in Figure \ref{fig:nil}, SAPBERT achieves lower accuracy (maximum value 0.7920) no matter what the threshold value is, which corresponds to our assumptions. When the threshold value is low, SAPBERT generates many wrong predictions to unlinkable query concepts; otherwise, SAPBERT continues to output ``NIL'' for many linkable concepts.

\subsection{Case Studies}
In case studies on linking EHR concepts to \datasetone's KG disease concepts, three scenarios are presented: (1) concepts assessed by both ground-truth labels and a clinician; (2) concepts evaluated by a clinician due to missing ground-truth labels; (3) irrelevant concepts judged by a clinician. The linking results of \modelname\ and SAPBERT are presented in Table \ref{tab:cases}. 
Overall, \modelname\ could link biomedical concepts more accurately and appropriately.
For casse I-V, \modelname's linking results are justified by the ground-truth label and clinician. Specifically, for cases I and II, \modelname\ accurately links the EHR concepts to conceptually similar
but lexically distinct KG concepts, while SAPBERT links to lexically similar but conceptually different KG concepts. This difference showcases the effective use of LLM's biomedical knowledge. SAPBERT also shows inaccuracies in cases III-IV, and provides a broader prediction in case V, whereas \modelname's predictions are more accurate and specific. 
For cases VI-IX, where linking ground truth labels are lacking, \modelname's predictions also align more accurately with EHR concepts than SAPBERT's, according to a clinician's review. In cases VI and VII, \modelname\ closely matches the EHR concepts, while SAPBERT's predictions are overly specific. In cases VIII and IX, \modelname\ correctly and automatically identifies no matching KG disease concepts, while SAPBERT fails to resolve that NIL prediction challenge unless manual thresholds are set and adjusted.

\begin{table}[ht]
    \caption{Analyzed cases. }
    \label{tab:cases}
    \centering
    \resizebox{\linewidth}{!}{%
    \begin{tabular}{cccc}
    \hline
       ID & EHR Concept & \modelname's Prediction & SAPBERT's Prediction\\
    \hline
       I & Chondroectodermal dysplasia & Ellis-van Creveld syndrome \humanEmoji \labelEmoji & Cranioectodermal dysplasia \\
       II & Dermatophytosis of hand & Tinea manuum \humanEmoji \labelEmoji & Hand dermatosis \\
       III & Late syphilis, unspecified & Tertiary syphilis \humanEmoji \labelEmoji & Secondary syphilis \\
       IV &Hypopotassemia & Hypokalemia \humanEmoji \labelEmoji & Hypocupremia nos \\
       V &Epidemic vertigo & Vestibular neuronitis \humanEmoji \labelEmoji & Vertigo \\
    \hline
       VI & Postprocedural fever & Postoperative complications \humanEmoji & Postcardiotomy syndrome  \\
       VII & Acquired cardiac septal defect & Heart septal defect \humanEmoji & Atrial heart septal defect\\
     \hline
       VIII & Height of bed & NIL \humanEmoji & Binge eating disorder \\
       IX & Level one & NIL \humanEmoji & Glaucoma 1 open angle\\
    \hline
    \end{tabular}%
    }
    \begin{flushleft}
    \scriptsize
    Note: ``\humanEmoji'' indicates this prediction is justified by the clinician. ``\labelEmoji'' indicates this prediction is justified by the ground-truth label.
    \end{flushleft}
\end{table}

%% file: sections/6_conclusion.tex
\section{Conclusion}

In this study, we introduce \modelname, a novel framework leveraging LLMs and multi-stage prompts for effective biomedical concept linking. 
Compared with previous concept linking methods, \modelname\ achieves better linking accuracy, attributed to LLM's intrinsic strong biomedical knowledge. \modelname\ further employs multi-stage prompts to maintain cost-efficiency and handle the NIL prediction problem. Moreover, \modelname\ functions as a zero-shot framework, requiring no training and demonstrating strong flexibility and generalizability across biomedical systems. 
Promising future work can focus on further enhancing the prompt effectiveness, reducing costs, and minimizing manual efforts, aiming to extend \modelname's application to broader systems.